\documentclass[%
reprint,
amsmath,amssymb,
aps,
]{revtex4-1}
\usepackage{graphicx}% Include figure files
\usepackage{dcolumn}% Align table columns on decimal point
\usepackage{bm}% bold math
\usepackage{hyperref}% add hypertext capabilities
\usepackage{color}
\newcommand{\beq}{\begin{equation}}
\newcommand{\eeq}{\end{equation}}
\newcommand{\bea}{\begin{eqnarray}}
\newcommand{\eea}{\end{eqnarray}}

\begin{document}

%\preprint{APS/123-QED}

\title{Shear viscosity $\eta$ to electric conductivity $\sigma_{el}$ ratio for the Quark-Gluon Plasma}
%\thanks{A footnote to the article title}%

\author{A. Puglisi}
% \altaffiliation[Also at ]{Physics Department, XYZ University.}%Lines break automatically or can be forced with \\
%\email{puglisia@lns.infn.it}
\author{S. Plumari}%
% \email{salvatore.plumari@ct.infn.it}

\author{V. Greco}
%\email{greco@lns.infn.it}
\affiliation{Department of Physics and Astronomy, University of Catania, Via S. Sofia 64, I-95125 Catania, Italy}
\affiliation{Laboratorio Nazionale del Sud, INFN-LNS, Via S. Sofia 63, I-95125 Catania, Italy }

\date{\today}

\begin{abstract}
The transport coefficients of strongly interacting matter are currently subject of intense theoretical and phenomenological studies due to their
relevance for the characterization of the quark-gluon plasma produced in ultra-relativistic heavy-ion collisions (uRHIC).
We discuss the connection between the shear viscosity to entropy density ratio, $\eta/s$, and the electric conductivity, $\sigma_{el}$. 
Once the relaxation time is tuned to have a minimum value of $\eta/s=1/4\pi$ 
near the critical temperature  $T_c$, one simultaneously predicts $\sigma_{el}/T$ very close to recent lQCD data.
More generally, we discuss why the ratio of $(\eta/s)/(\sigma_{el}/T)$ supplies a measure of the quark to gluon scattering rates whose knowledge 
would allow to significantly advance in the understanding of the QGP phase.
We also predict that $(\eta/s)/(\sigma_{el}/T)$, independently on the running coupling $\alpha_s(T)$, 
should increase up to about $\sim 20$ for $T \rightarrow T_c$, while it goes down to a nearly
flat behavior around $\simeq 4$ for $T \geq 4\, T_c$. Therefore we in general predict a stronger T dependence of $\sigma_{el}/T$ with respect to $\eta/s$ that in a quasiparticle approach is constrained by lQCD thermodynamics. A conformal theory, instead, predicts a similar T dependence of $\eta/s$ and $\sigma_{el}/T$.
\end{abstract}


\maketitle

Relativistic Heavy Ion Collider (RHIC) at BNL and Large Hadron Collider (LHC) at CERN have produced a very hot and dense system of strongly interacting particles as in the Early Universe with temperatures largely above $T_c\simeq 160 \rm MeV$  \cite{Science_Muller,Shuryak:2003xe,Aoki:2006we},
the transition temperature from nuclear matter to the Quark-Gluon Plasma (QGP) \cite{lQCD,Bazavov:2011nk,Lombardo:2012ix}.
The phenomenological studies by viscous hydrodynamics \cite{Romatschke:2007mq,Song:2011hk,Schenke:2010nt,Niemi:2011ix}
and parton transport \cite{Ferini:2008he,Plumari_Bari,Xu:2007jv,Xu:2008av,Cassing:2009vt,Bratkovskaya:2011wp} of the collective behavior
have shown that the QGP has a very $\eta/s$, quite close to the conjectured lower-bound limit  for a strongly interacting system in the limit of infinite coupling $\eta/s=1/4\pi$ \cite{Kovtun:2004de}.
This suggests that hot QCD matter could be a nearly perfect fluid with the smallest $\eta/s$ ever observed, even less dissipative than the ultra cold matter created by magnetic traps \cite{O'Hara:2002zz,Cao:2010wa}.
As for atomic and molecular systems a minimum in $\eta/s$ is expected slightly above $T_c$ \cite{Csernai:2006zz,Lacey:2006bc}.

Another key transport coefficient, yet much less studied, is $\sigma_{el}$.
This transport coefficient represents the linear response of the system to an applied external electric field. 
Several processes occurring in uRHIC as well as in the Early Universe are regulated by the electric conductivity. Indeed
HICs are expected to generate very high electric and magnetic fields 
($eE\simeq eB \simeq m_{\pi}^2$, with $m_{\pi}$ the pion mass) in the very early stage of the collisions \cite{Tuchin,Hirono}. 
A large value of $\sigma_{el}$ would determines
a relaxation time for the electromagnetic field of the order of $\sim 1-2\, fm/c $ \cite{McLerran:2013hla,Gursoy:2014aka}, which would be
of fundamental importance for the strength of the Chiral-Magnetic Effect \cite{Fukushima:2008xe}, a signature of  the CP violation 
of the strong interaction. Also in mass asymmetric collisions, like Cu+Au, the 
electric field directed from Au to Cu induces a current resulting in charge asymmetric
collective flow directly related to $\sigma_{el}$ \cite{Hirono}.
Furthermore the emission rate of soft photons should be directly proportional to $\sigma_{el}$
\cite{Kapusta_book,Turbide:2003si,Linnyk:2013wma}.
Despite its relevance there is yet only a poor theoretical and phenomenological knowledge of $\sigma_{el}$ and its temperature
dependence. First preliminary studies in lQCD has extracted only few estimates with large uncertainties \cite{Gupta, Aarts} and only
recently more safe extrapolation has been developed \cite{Amato,Ding,Brandt}.

In this Letter, we point out the main elements determining $\sigma_{el}$ for a QGP plasma and in particular its
connection with $\eta$. In fact, while one may expect that the QGP is quite a good conductor 
due to the deconfinement of color charges, on the other hand, the very small $\eta/s$ indicates large scattering rates which can largely damp the conductivity, especially if the plasma is dominated by gluons that do not carry any electric charge.

The electric conductivity can be formally derived from the Green-Kubo formula and it is related to the relaxation
of the current-current correlator for a system in thermal equilibrium. It can be written as $\sigma_{el}=V/(3\,T)\, \langle \vec J (t=0)\cdot \vec J (t=0)\rangle \cdot \tau$, where $\tau$
is the relaxation time of the correlator whose initial value can be related to the thermal
average $\frac{ \rho\, e^2}{3T}\langle p^2/E^2 \rangle $ \cite{FernandezFraile:2005ka}, with $\rho$
and $E$ the density and energy of the charge carriers.
Generalizing to the case of QGP one can write:
\begin{equation}
\sigma_{el}=\frac{e^2}{3T} \left \langle \frac{{\vec p}^{\,2}}{E^2}\right\rangle \sum_{j=q,\bar q} f_j^2 \,\tau_j \rho_j=
\frac{e_\star^2}{3T} \left \langle \frac{{\vec p}^{\,2}}{E^2}\right\rangle \tau_q \rho_q
\label{conductivity_qgp}
\end{equation}
where $e_\star^2=e^2\sum_{j=u,d,s}^{\bar{u},\bar{d},\bar{s}}f_j^2=4e^2/3  $ with $f_j$ the fractional quark charge. 
Eq. (\ref{conductivity_qgp}) in the non-relativistic limit reduces formally to the Drude formula $\frac{\tau e^2 \rho}{m}$, even if
we notice that $\tau$ in Eq.(\ref{conductivity_qgp}) has not to be equal to $1/(\sigma\rho)$ as in the Drude model. 
The relaxation time of a particle of species $j$ in terms of cross-sections and particle densities can be written in the relaxation
time approximation (RTA) as 
$ \tau_j^{-1}=\sum_{i=q, \bar q , g} \langle \rho_i v_{rel}^{ij} \sigma^{ij}_{tr}\rangle$
where $j=q, \bar q$ while the  sum runs over all particle species with $\rho_i$ the density of species $i$, $v_{rel}^{ij}$ is the relative velocity and $\sigma^{ij}_{tr}$ is the transport scattering cross-section. 
%\textcolor{blue}{This definition of $\tau$ is the one used in all our computations and results.}
In Ref. \cite{Cassing_el} it has been shown that RTA is able to describe 
with quite good approximation $\sigma_{el}$ in agreement with numerically simulation of the Dynamical QP model (DQPM) known as PHSD, see also more generally for a numerical approach Ref.s \cite{Greif:2014oia,Puglisi:2014sha}. 

As done within the Hard-Thermal-Loop (HTL) approach, we will consider the total transport cross section regulated by a screening Debye mass  $m_D=g(T)T$, with $g(T)$ being the strong coupling:
\begin{equation}\label{pQCD_cross-section}
\sigma_{tr}^{ij}(s)=\int \frac{d\sigma}{dt} \sin^2 \Theta\,dt=\beta^{ij}\frac{\pi \alpha^2_s}{m_D^2}\frac{s}{s+m_D^2}h(a)
\end{equation}
where $\alpha_s=g^2/4\pi$, the differential cross section $\frac{d\sigma}{d t}=\frac{d\sigma}{d q^2}\simeq \alpha_s^2/(q^2+m_D^2)^2$ where $q^2=\frac{s}{2}(1-\cos \theta)$. The function $h(a)=4a(1+a)[(2a+1)\ln(1+1/a)-2]$, with $a=m_D^2/s$ accounts for the anisotropy of the scatterings: for $m_D\to \infty$, $h(a)\to 2/3$ and one recovers the isotropic limit 
%while $h(a)<2/3$ for finite value of $m_D$.
The coefficient $\beta^{ij}$ depends on the pair of interacting particles:
$\beta^{qq}=16/9$, $\beta^{qq'}=8/9$, $\beta^{qg}=2$, $\beta^{gg}=9$. 
These factors are directly related to the quark and gluon Casimir factor, for example $\beta^{qq}/\beta^{gg}=(C_F/C_A)^2=(4/9)^2$.

The shear viscosity $\eta$
is known from the Green-Kubo relation to be given by $\eta=V/T\, \langle \Pi_{xy}^2(t=0)\rangle\,\cdot \tau$,
where the initial value of the correlator of the transverse components of the energy-momentum
tensor can be written as $\frac{\rho}{15 T}  \langle p^4/E^2 \rangle$
\cite{Plumari_visco,Wesp:2011yy,Fuini:2010xz}. 
Hence for a system with different species can be written as \cite{Sasaki,Kapusta_qp}:
\begin{equation}
\eta=\frac{1}{15 T}  \left\langle \frac{p^4}{E^2} \right\rangle \left(  \tau_q \rho^{tot}_q  +  \tau_g \rho_{g}\right)
\label{viscosity}
\end{equation}
where the relaxation time $\tau_g$ has a similar expression as above with $j=g$ while $\rho^{tot}$ is the sum of all quarks and antiquarks flavour density.
The thermodynamical averages entering Eq.s (\ref{conductivity_qgp}) and (\ref{viscosity}), 
will be fixed employing
a quasi-particle (QP) model tuned to reproduce the lattice QCD thermodynamics \cite{Plumari_qpmodel}, similarly to \cite{Levai:1997yx,Peshier:2002ww,Bluhm:2010qf,Bluhm:2004xn}. 
%The QP model describes a strongly interacting system in terms of weakly interacting particles whose masses $m_{q,g}(T)$ arise from the non-perturbative interaction and account for most of it, while residual interaction is treated in terms of a background field known as bag constant.
%Hence the system can be treated as a gas of T-dependent massive particles under the action of a background field.
The quark and gluon masses are given by $m^2_g= 3/4\,g^2 T^2$ and $m^2_q=1/3\,g^2 T^2$
in terms of a running coupling $g(T)$ that is determined by a fit to the lattice energy density, 
which allows to well describe also the pressure $P$ and entropy density $s$ above $T_c=160 \, \rm MeV$. 
In Ref. \cite{Plumari_qpmodel} we have obtained:
\begin{equation}
g^2_{QP}(T)={48 \pi^2}/{\left( 11 N_c - 2 N_f\right) \ln\left[ \lambda \left( \frac{T}{T_c}-\frac{T_s}{T_c} \right) \right]^2}
\end{equation}
with $\lambda = 2.6$, $T_s/T_c=0.57$. We warn that the previous equation is a good parametrization only for $T>1.1\,T_c$.
We notice that a self-consistent dynamical model (DQPM),
that includes also the pertinent spectral function, has been developed in \cite{Cassing:2009vt} and leads to nearly the same 
behavior of the strong coupling $g(T)$. 
%However, the simple QP model has the advantage
%to handle simpler analytical expression to pin down the core physics.
We will consider the DQPM explicitly, showing that the considerations elaborated in this Letter
are quite general and can be only marginally affected by particle width.
%We notice that 
%the correct thermal averages entering the transport coefficients, Eq. (\ref{viscosity}) and Eq (\ref{conductivity_qgp}), have been determined fitting $g(T)$ to lQCD thermodynamics, but this does not imply that with the same $g(T)$ one has the correct description also of the scattering dynamics,
%unless one believes in QP model as a solid microscopic description of the correct one, which is not necessarily our point of view. 

We notice that the only approximation made in deriving Eq.(\ref{viscosity}) is to consider
$\langle p^4/E^2\rangle$ equal for quarks and gluons.
We have verified that $\langle p^4/E^2 \rangle_{g}\simeq \langle  p^4/E^2 \rangle_q$ within a 5$\%$ in the QP model but also more generally even when
$m_q$ and $m_g$ are largely different but $m_{q,g} \lesssim 3T$, which means that Eq.(3) is valid also for light and strange current quark masses and massless gluons. 
The $\langle p^4/E^2\rangle$ in a massless approximation is simply $4\epsilon\, T /\rho$,
we have checked that the validity of this expression is kept using the QP model (i.e. massive excitation) with a discrepancy of about $2\%$.
Hence the first term in Eq. (\ref{viscosity}) is determined by the lQCD thermodynamics and does not rely on the detailed $m_{q,g}(T)$ in the QP model.
We note that even if the QP model is able to correctly describe 
the thermodynamics it is not obvious that  it correctly describes dynamical quantities like the relaxation times with the same coupling $g(T)$ employed to fit  the thermodynamics. However our key point will be to find a quantity independent on $g(T)$, see Eq.(5).

For its general interest and asymptotic validity for $T\rightarrow \infty$,
we also consider the behavior of the pQCD running coupling constant for the evaluation of transport relaxation time:
$g_{pQCD}(T)=\frac{8\pi^2}{9} \ln^{-1} \left(\frac{2\pi T}{\Lambda_{QCD}} \right)$.
On one hand, close to $T_c$, such a case misses the dynamics of the phase transition,
on the other hand it allows to see explicitly what is the impact of a
different running coupling.
\begin{figure}
 \centering
 \includegraphics[scale=0.28, keepaspectratio=true]{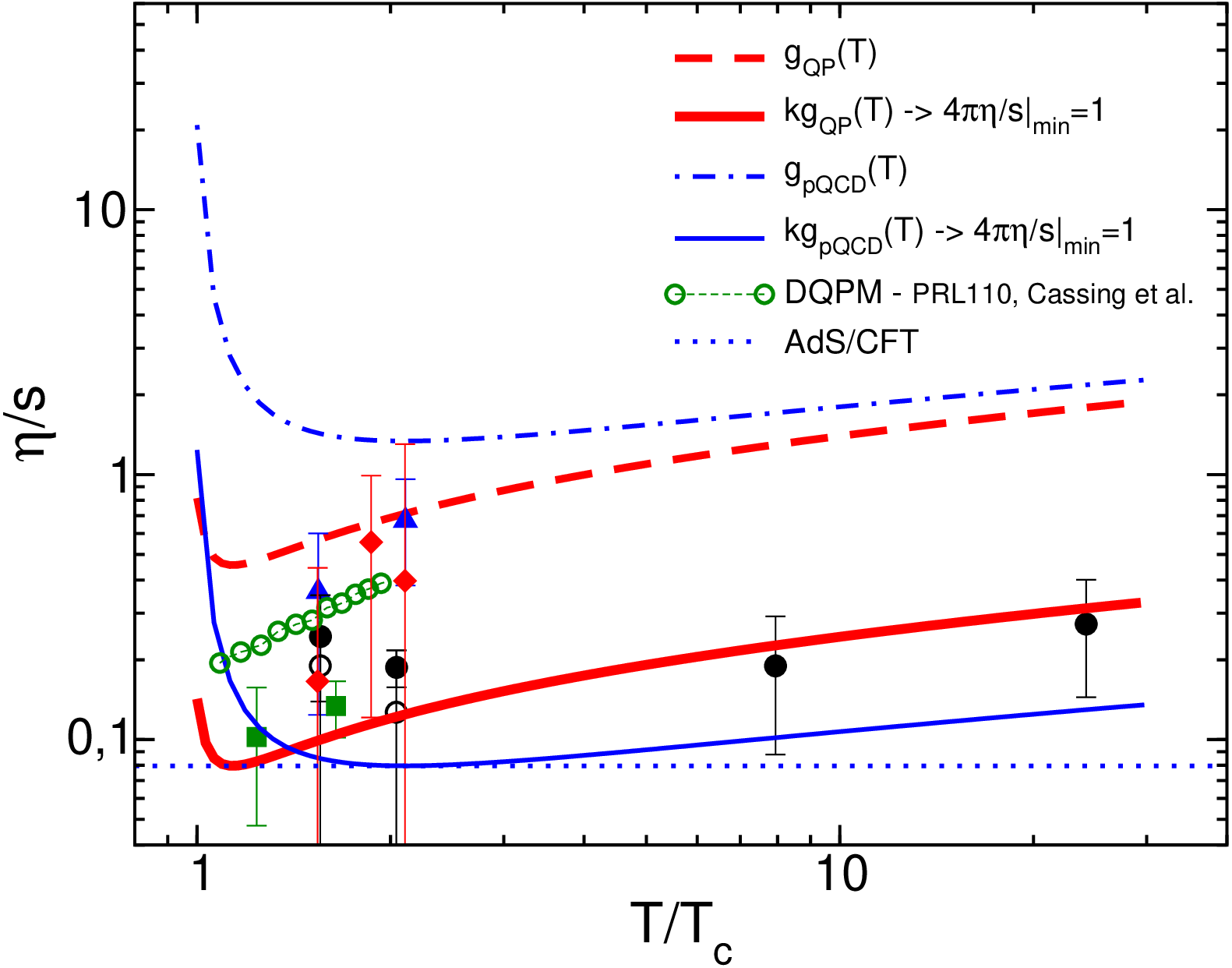}
 \caption{Shear viscosity to entropy density ratio $\eta/s$: dashed line represents QP model results, dot-dashed line is pQCD, stars is DQPM \cite{Marty}. Red thick solid line and blue thin solid line are obtained rescaling $g(T)$. Blue dotted line is AdS/CFT result from \cite{Kovtun:2004de}.
 Symbols are lattice date: full squares \cite{Meyer}, diamonds and triangles \cite{Nakamura}, open and full circles \cite{Sakai}.}
 \label{fig:eta_s_Lattice}
\end{figure}
The $\eta/s$ calculated is shown 
in Fig. \ref{fig:eta_s_Lattice}: red dashed line is the result for the QP model 
using $g_{QP}(T)$ for relaxation times and transport coefficient, blue dot-dashed line labeled as $g_{pQCD}$, means that we used the pQCD running coupling for evaluating
the relaxation time, 
green stars are the DQPM \cite{Marty} and by symbols several lQCD results. We warn that the different lQCD data are obtained with different methods and actions.
The main difference between our QP model and DQPM comes from the fact that the latter assumes 
isotropic scatterings which decrease the relaxation time by about $30-40\%$.
%We have checked that for isotropic scatterings the results become almost identical.
Anyway, the $\eta/s$ predicted is toward higher value with respect to
the conjectured minimum value of $\eta/s\sim 0.08$, supported also by several phenomenological estimates \cite{Romatschke:2007mq,
Song:2011hk,Schenke:2010nt,Niemi:2011ix,Ferini:2008he}.
%This means that if the $\eta/s$ is really very close to $1/4\pi$ then the relaxation time estimated is still too large. 
However within the QP model it has been discussed in the literature also another approach for $\tau$ where the relaxation times are $\tau_{q,g} = C_{q,g} \,g^4 T \ln (a/g^2)$ \cite{Khvorostukhin} with $C_{q,g}$ and $a$ fixed to reproduce both
the pQCD estimate asymptotically \cite{Arnold:2003} and a minimum for  $\eta/s(T)=1/4\pi$ \cite{Plumari_qpmodel,Bluhm:2010qf}.
In the T region of interest, the result is quite similar
to upscaling the coupling $g(T)$ by a $k$-factor in such a way to have the minimum of $\eta/s(T)=1/4\pi$.
Therefore we do not employ the above parametrization but compute the transport coefficients using the definition of $\tau$ of Eq.(\ref{tau_transport}), where enters the cross section in Eq. (\ref{pQCD_cross-section}) with the coupling upscaled.
The corresponding curves are shown in Fig. \ref{fig:eta_s_Lattice} by red thick solid line for the $g_{QP}(T)$ coupling (rescaled by $k=1.59$)
and by blue thin solid line for the $g_{pQCD}(T)$ (rescaled by $k=2.08$). One obtains $\tau_g \simeq \tau_q/2 \sim 0.2 \rm\, fm/c$ and
also $\eta/s(T)$ roughly linearly rising with $T$
in agreement with quenched lQCD estimates, full circles \cite{Sakai}.

\begin{figure}
 \centering
 \includegraphics[scale=0.28, keepaspectratio=true]{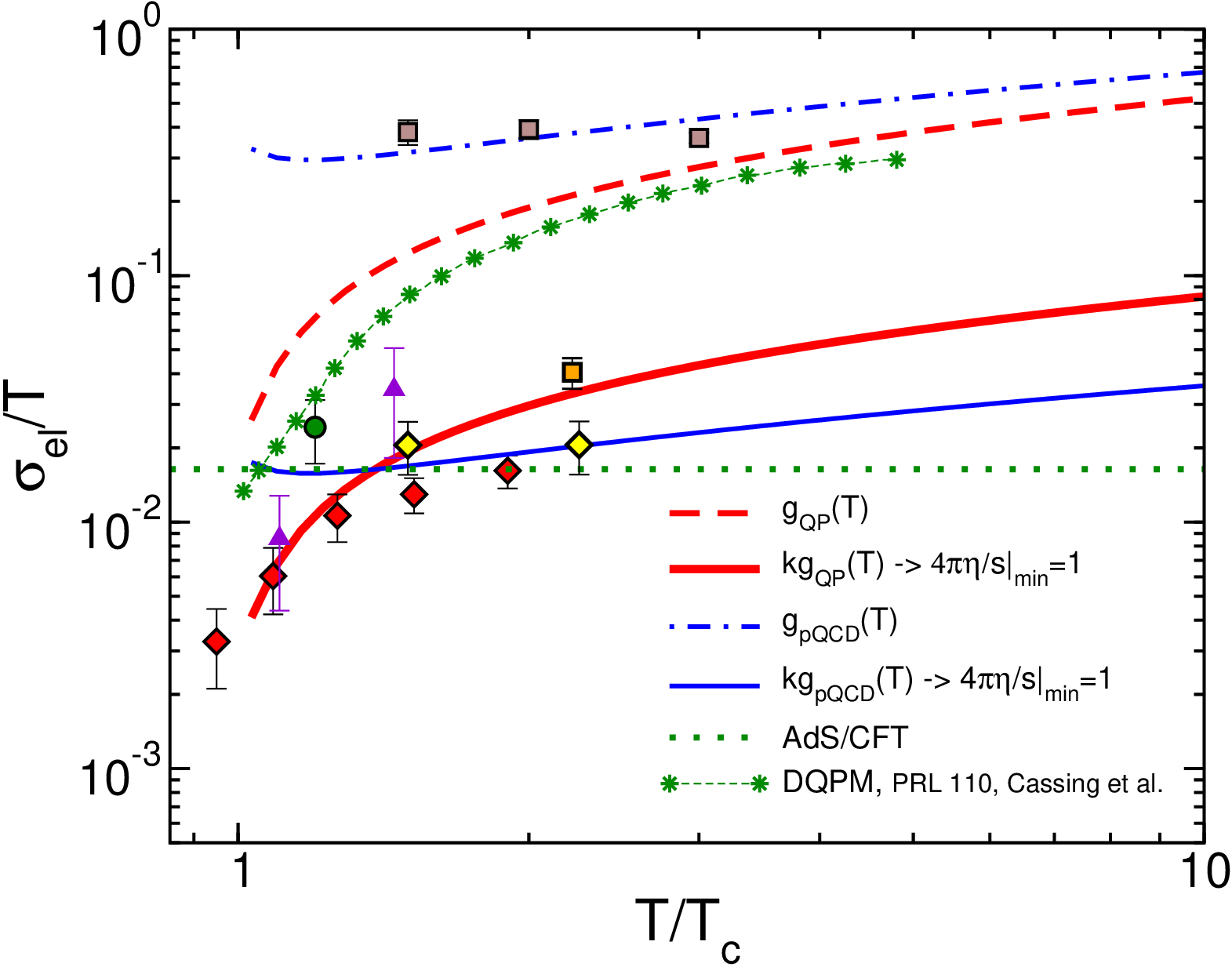}
 \caption{Electric conductivity $\sigma_{el}/T$ as a function of $T/T_c$: red dashed line represents QP model results, blue dot-dashed line is pQCD, red thick solid line and blue thin solid line are respectively QP and pQCD considering the rescaled $g(T)$ in order to reproduce the minimum of $\eta/s$. Green line are AdS/CFT results from \cite{Huot}.  Green stars represent  DQPM \cite{Cassing_el}. Symbols are Lattice data: grey squares \cite{Gupta}, violet  triangles \cite{Ding}, green circle \cite{Brandt}, yellow diamond \cite{Aarts}, orange square\cite{Buividovich:2010tn} and red diamonds \cite{Amato}.}
 \label{fig:K_el_cond}
\end{figure}

A main point we want to stress is that,
once the relaxation time is set to an $\eta/s(T)=0.08$, the $\sigma_{el}/T$ predicted, with the same $\tau_q$ as for $\eta/s$, is in quite good agreement
with most of the lQCD data, shown by symbols in Fig. \ref{fig:K_el_cond} (see caption for details).
Therefore a low $\sigma_{el}/T$  is obtained at variance with the early lQCD estimate, Ref. \cite{Gupta},
as a consequence of the small $\tau_{q,g}$ entailed by $\eta/s \simeq 0.08$. 
In Fig. \ref{fig:K_el_cond}, we show also the predictions of DQPM (green stars) \cite{Cassing_el, Marty}.

In Fig.\ref{fig:K_el_cond}, we also plot by green dotted line the ${\cal N}=4$ Super Yang Mills  electric conductivity \cite{Huot}
that predicts a constant behavior for $\sigma_{el}/T=e^2 N_c^2 / (16 \pi)$.
We note that in our framework one instead expects that, even if the $\eta/s$ is independent on the temperature,
the $\sigma_{el}$ should still have a strong T-dependence.
This can be seen noticing that
one can write approximately,  $\eta/s \simeq T^{-2} \tau \rho$, being $\langle p^4/E^2\rangle \simeq \epsilon T/\rho$,
and $\sigma_{el}/T\simeq T/m(T) \eta/s$, being $\langle p^2/E^2 \rangle \simeq T/m(T)$, which means an extra $T$ dependence for $\sigma_{el}$ leading to a steep decrease of $\sigma_{el}/T$ close to $T_c$. $m(T)$ increases as $T\to T_c$ because it is fitted to reproduce the decrease of energy density $\epsilon$ in lQCD. We notice that for a conformal theory $T^{\mu}_{\mu}=\epsilon - 3P=0$, as for massless particles, one has $\sigma_{el}/T\sim \eta/s$ like found in AdS/CFT. It seems that the large interaction measure is the origin of such extra T dependence of $\sigma_{el}/T$ with respect to $\eta/s$. This indication is corroborated also by the recent result in AdS/QCD \cite{Noronha} that presents a similar strong T dependence for $T<2-3 T_c$ at variance with AdS/CFT.

The $\sigma_{el}$ appears to be self-consistent with a minimal $\eta/s$, but the
specific $T$ dependence of both are largely dependent on the modeling of $\tau_{q,g}$,
we point out that the ratio $(\eta/s)/(\sigma_{el}/T)$ can be written, 
from Eq. (\ref{conductivity_qgp}) and Eq. (\ref{viscosity}), as:
\begin{align}\label{ratio}
\frac{\eta/s}{\sigma_{el}/T}=\frac{6}{5}\frac{ T \langle p^2/E^2 \rangle^{-1}}{s\,e_\star^2 } \left\langle \frac{p^4}{E^2} \right\rangle \left( 1+ \frac{\tau_g}{\tau_q} \frac{\rho_g}{\rho^{tot}_{q}}  \right).
\end{align}
in terms of generic relaxation times. Eq.(\ref{ratio}) is quite general and does not rely on specific features or validity of the quasi-particle model.
A main feature of such a ratio is its independence on the $k$-factor introduced above,
and, more importantly, even on the $g(T)$ coupling as we can see writing explicitly the transport relaxation time for quarks and gluons:
\begin{eqnarray}\label{tau_transport}
\tau_q^{-1} =\langle \sigma(s)_{tr} v_{rel} \rangle (\rho_q \sum_{i=u,d,s}^{\bar{u},\bar{d},\bar{s}} \beta^{qi} + \rho_g \beta^{qg}) \nonumber\\
\tau_g^{-1}= \langle \sigma(s)_{tr} v_{rel} \rangle \left(\rho_q^{tot} \beta^{qg} + \rho_g\beta^{gg}\right)
\end{eqnarray}
where the $\beta^{ij}$ were defined above. Hence
the ratio of transport relaxation times appearing in Eq. (\ref{ratio}) can be written as:
\begin{align}
\label{tau-ratio}
\frac{\tau_g}{\tau_q} =  \frac{{C^q}+ \frac{\rho_g}{\rho_q}}{6 + \frac{\rho_g}{\rho_q} C^g }
\end{align}
where the coefficients $C^{q}=(\beta^{qq} + \beta^{q\bar{q}} + 2 \beta^{q\bar{q}'} + 2 \beta^{qq'})/\beta^{qg}$ and $C^{g}=\beta^{gg}/\beta^{qg}$
are the relative magnitude between quark-(anti-)quark and $gg$ with respect to $q(\bar q)g$ scatterings.
\begin{figure}
 \centering
 \includegraphics[scale=0.28, keepaspectratio=true]{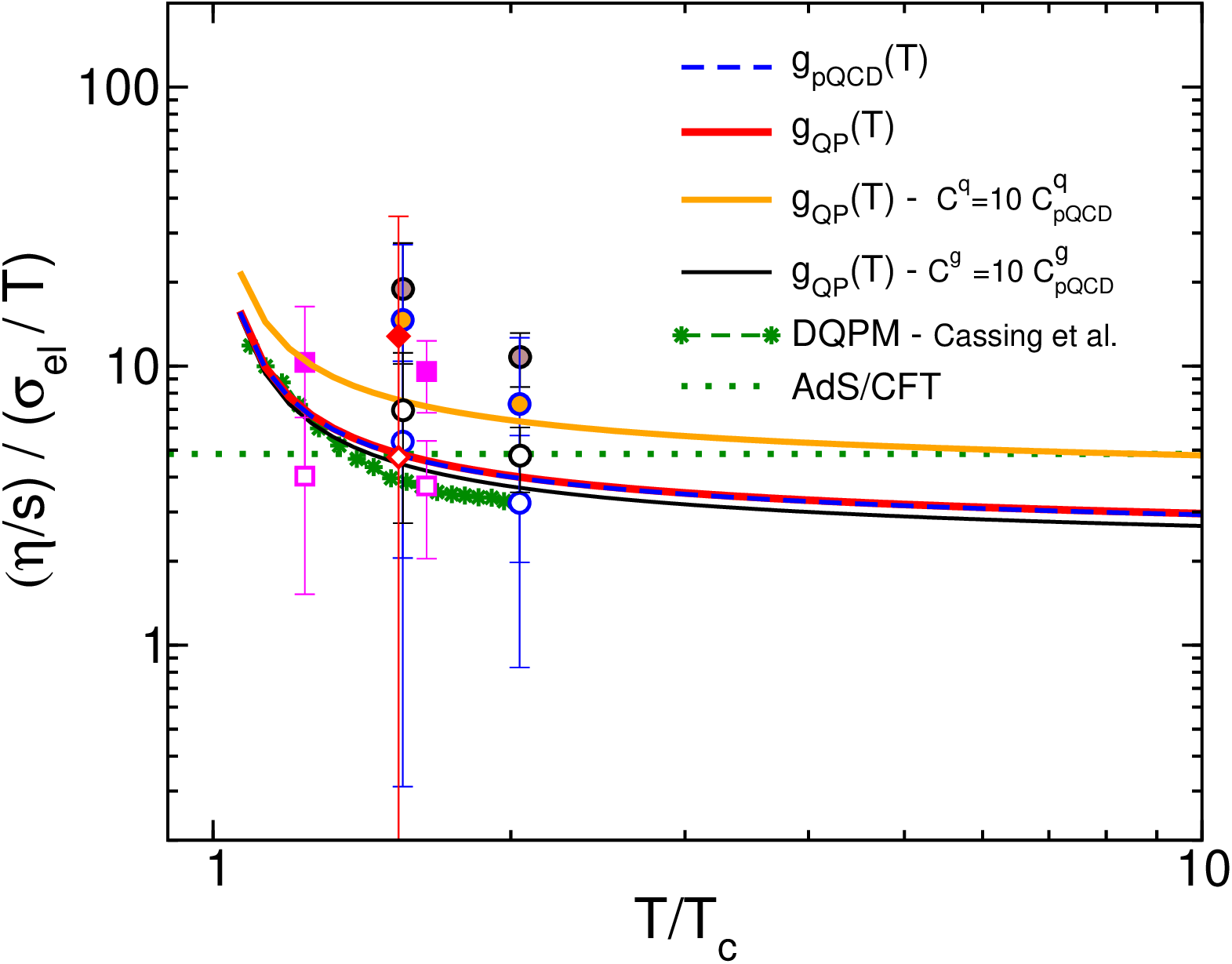}
 \caption{Shear viscosity $\eta/s$ to $\sigma/T$ ratio as a function of $T/T_c$: 
red solid line is the QP model, blue dashed line pQCD, green stars DQPM \cite{Marty}. 
Orange line is obtained using $C^q=10\,C^q_{pQCD}$, black thin line $C^g=10\,C^g_{pQCD}$.
Green dotted line represent AdS/CFT results \cite{Kovtun:2004de, Huot}.
Symbols are obtained using available lattice data (see text for details).}
 \label{fig:ratio}
\end{figure}
%The factor 6 expresses simply the number of flavors and anti-flavors.
Using the standard pQCD factors for $\beta_{ij}$,
$C^{q}|_{pQCD}=\frac{28}{9}\simeq3.1$ and $C^{g}|_{pQCD}=\frac{9}{2}$. 

In Fig. \ref{fig:ratio} we show $(\eta/s)/(\sigma_{el}/T)$ as a function of $T/T_c$: the red thick solid line is 
the prediction for the ratio using $g_{QP}(T)$, but it is clear
from the Eq. (\ref{ratio}) that the ratio is completely independent on the running coupling itself; the result for $g_{pQCD}(T)$ 
is shown by blue dashed line.
The ratio is instead sensitive just to the relative
strength of the quark (anti-quark) scatterings with respect to the gluonic ones, hence we suggest that
a measurement in lQCD can shed light on the relative scattering rates of quarks and gluons, providing an insight into their relative role.
It is not known if such ratios, linked to the Casimir factors of $SU(3)_c$, are kept also in the non-perturbative regime, which
may be not so unlikely \cite{Nakamura:2005hk}.
We remark that we have computed the ratio in a very large temperature range $1-10\,T_c$: at large temperatures ($T> 5-10\,T_c$) deviation
from the obtained value, $(\eta/s)/(\sigma_{el}/T)\simeq 3$, would be quite surprising, on the other hand 
for $T<1.2-1.5\, T_c$ one may cast doubts on the validity of the Casimir coefficients. In the following we evaluate also the impact of modified Casimir Coefficients.
As $T\rightarrow T_c$ a steep increase is predicted that is essentially regulated by $\langle p^2/E^2 \rangle$.
It is interesting to notice that in the massless limit (conformal theory) the factor before the parenthesis in Eq.(\ref{ratio}) becomes a temperature independent constant and hence also the ratio. This is in quite close 
agreement with the AdS/CFT prediction shown by dotted line in Fig. \ref{fig:ratio}.

We also briefly want to mention that one possible scenario could be that when the QGP
approaches the phase transition, the confinement dynamics becomes dominant and 
the $q\bar q$ scattering, precursors of mesonic states, and
di-quark $qq$ states, precursor of baryonic states, are strongly enhanced 
by a resonant scattering with respect to other channels, as found in a T-matrix
approach in the heavy quark sector \cite{vanHees:2007me}. 
For this reason, we explore the sensitivity of the ratio $(\eta/s)/(\sigma_{el}/T)$ on the magnitude of $C^{q}$ and $C^{g}$.
The orange solid line shows the behavior for an enhancement of the quark scatterings,
$C^q=10\,C^q_{pQCD}$. We can see in Fig. \ref{fig:ratio} that this would lead to an enhancement of the ratio by about a $40\%$.
We also see that instead the ratio is not very sensitive to a possible enhancement of only the $gg$ scattering with
respect to the $q\bar q,qq,qg$; in fact even for $C^g=10 C^g_{pQCD}$ one obtains the
thin black solid line.
This is due to the fact that already in the pQCD case $\tau_g/\tau_q \sim 0.3-0.4$. 
Furthermore already in the massless limit $\rho_g/\rho_q^{tot} \simeq d_g/d_{q+\bar q}=4/9$ even not dwelling on the details of the QP model where the larger gluon mass further decreases this ratio. Therefore
the second term in parenthesis in Eq. (\ref{ratio}) is of the order of $10^{-1}$ and further decrease of its value would not
be visible because the ratio is anyway dominated by the first term equal to one. 
%Also for this reason the $(\eta/s)/(\sigma_{el}/T)$
%should be a quite stable predictions and large deviation from the predicted value would be a signature of
%unknown and unexpected properties of the QGP.
We reported in Fig. \ref{fig:ratio} also the ratio from the DQPM model, as deduced from \cite{Marty} and we can see that, even if it is not evaluated through Eq. (\ref{ratio}), it is in very good agreement with our general prediction.
In Fig. \ref{fig:ratio} we also display by symbols the ratio evaluated from the available lQCD data, considering 
for $4\pi \eta/s \lesssim 4$ while for $\sigma_{el}/T$ we choose 
red diamonds \cite{Amato} as a lower limit (filled symbols) and the others in Fig. \ref{fig:K_el_cond} as an upper limit (open symbols), excluding only
the grey squares \cite{nota-gupta}.
To compute $(\eta/s)/(\sigma_{el}/T)$ 
we do an interpolation between the data point of $\sigma_{el}$.
We warn to consider these estimates
only as a first rough indications, in fact the lattice data collected are
obtained with different actions among them and have quite different $T_c$ with respect to the most realistic one, $T_c\sim 160\,MeV$ \cite{lQCD,Bazavov:2011nk}, that we employed to tune the QP model \cite{Plumari_qpmodel}.
%and also with respect to the most realistic one for the EoS \cite{lQCD,Bazavov:2011nk},
%that we employed to tune the QP model \cite{Plumari_qpmodel}.

In this Letter we point out the direct relation between the shear viscosity $\eta$ and the electric conductivity $\sigma_{el}$.
In particular, we have discussed why most recent lQCD data \cite{Amato, Ding,Brandt} predicting an electric conductivity 
$\sigma_{el}\simeq 10^{-2} T$ (for $T < 2\, T_c$) ,
appears to be consistent with a fluid at the minimal conjectured viscosity $4\pi\eta/s\simeq 1$,
while the data of Ref. \cite{Gupta} appear to be hardly reconcilable with it.
Also a steep rise of $\sigma_{el}/T$, in agreement with lQCD data, appears quite naturally in the quasi-particle approach as inverse of the self-energy determining the effective masses needed to correctly reproduce the 
lQCD thermodynamics.
This result is at variance with the AdS/CFT \cite{Huot}, but our analysis suggests that it is due to the conformal thermodynamics that does not reflect the QCD one. 
It is quite interesting that an AdS/QCD approach \cite{Noronha}, able to correctly 
describe the interaction measure of lQCD, also modify the AdS/CFT result predicting a strong T dependence of $\sigma_{el}/T$ for $T<2-3\,T_c$. We note that the extra T dependence predicted for $\sigma_{el}/T$ with respect to $\eta/s$ is determined by the $\langle p^2/E^2 \rangle$ constrained to reproduce the lQCD thermodynamics. If instead one imposes conformality with $m=0$, this leads to $\langle p^2/E^2 \rangle=1$ and the T dependence of $\eta/s$ becomes quite similar to the one of $\sigma_{el}/T$ apart from differences that can arise between quark and gluon relaxation times. 

We identify the dimensionless ratio $(\eta/s)/(\sigma_{el}/T)$ as not affected by the uncertainties in the
running coupling $g(T)$. Moreover due to the fact that gluons do not carry an electric charge, 
the ratio is regulated by the relative strength and chemical composition of the QGP through the term $(1+ \tau_g \rho_g/\tau_q \rho_q^{tot})$.
Our analysis provides the baseline of such a ratio that in this decade will most likely be more safely evaluated
thanks to the developments of lQCD techniques. This will provide a first and pivotal insight into the understanding
of the relative role of quarks and gluons in the QGP. Deviations from our predictions for $(\eta/s)/(\sigma_{el}/T)$ especially at high temperature $T\gtrsim 2-3\,T_c$, where a quasi-particle picture can be derived 
from QCD within the HTL scheme \cite{Andersen:2010wu}, would be quite compelling.

\acknowledgments
V.G. acknowledge the support of the ERC-StG Grant under the QGPDyn project.
We thanks M. Ruggieri for carefully reading the manuscript.


\begin{thebibliography}{99}
\bibitem{nota-gupta}
These data are not considered being the only set more then one order of magnitude above all the other lQCD data.
\end{thebibliography}
\end{document}